\documentclass[twocolumn,nopacs,preprintnumbers,amsmath,amssymb,pra]{revtex4}
\usepackage{color}
\usepackage{bm, graphicx, amsmath}
\usepackage{hyperref}

\newcommand{\graph}{\mathcal{G}}
\newcommand{\hi}{\mathcal{H}}

\begin{document}

\title{Finding paths with quantum walks or quantum walking through a maze}
\author{Daniel Reitzner} 
\affiliation{RCQI, Institute of Physics, Slovak Academy of Sciences, D\'ubravsk\'a cesta 9, 845 11 Bratislava, Slovakia}
\author{Mark Hillery}
\affiliation{Department of Physics, Hunter College of the City University of New York, 695 Park Avenue, New York, NY 10065 USA}
\affiliation{Physics Program, Graduate Center of the City University of New York, 365 Fifth Avenue, New York, NY 10016}
\author{Daniel Koch}
\affiliation{Department of Physics, Hunter College of the City University of New York, 695 Park Avenue, New York, NY 10065 USA}
\affiliation{Physics Program, Graduate Center of the City University of New York, 365 Fifth Avenue, New York, NY 10016}

\begin{abstract}
We show that it is possible to use a quantum walk to find a path from one marked vertex to another. In the specific case of $M$ stars connected in a chain, one can find the path from the first star to the last one in $O(M\sqrt{N})$ steps, where $N$ is the number of spokes of each star. First we provide an analytical result showing that by starting in a phase-modulated highly superposed initial state we can find the path in $O(M\sqrt{N}\log M)$ steps.  Next, we improve this efficiency by showing that the recovery of the path can also be performed by a series of successive searches when we start at the last known position and search for the next connection in $O(\sqrt{N})$ steps leading to the overall efficiency of $O(M\sqrt{N})$. For this result we use the analytical solution that can be obtained for a ring of stars of double the length of the chain.
\end{abstract}

\maketitle

\section{Introduction}

Quantum walks are quantum versions of random walks \cite{davidovich,aharonov} (for reviews see \cite{reitzner,manoucheri}).  There are both discrete- and continuous-time versions of quantum walks \cite{farhi}, but here we will only make use of the discrete-time version.  There are also two (equivalent \cite{AnLu09,reitzner}) versions of the discrete-time walk, the coined walk and the scattering walk, and here we shall employ the scattering walk \cite{hillery3}, which is simple to use when working with non-regular graphs.

Quantum walks have proven useful in the development of quantum algorithms, particularly search algorithms \cite{shenvi,potocek,grid1,grid2,complete2,hillery1,kendon2,lee,feldman,hillery2,cottrell1,cottrell2}.  Originally, the searches were for marked vertices \cite{shenvi}, but it was later realized that searches for more general objects are possible.  One can search for marked edges or cliques \cite{hillery1}, extra edges that break the symmetry of a graph \cite{feldman}, or more general structures \cite{cottrell1,cottrell2}.  In addition to their use as theoretical tools, it has been possible to realize quantum walks in the laboratory \cite{PeLaPoSoMoSi08,ScMaScGlEnHuSc09,KaFoChStAlMeWi09,ScCaPo09,peruzzo,schreiber}.

A more recent use of quantum walks is in state transfer \cite{hein,stefanak,stefanak2}.  One has two distinguished vertices in a graph.  The particle making the walk starts on one, and the objective is for it to finish, after a certain number of steps, on the other with high probability.  This has been studied for grids \cite{hein}, star graphs and complete graphs with loops \cite{stefanak}, and complete bipartite graphs \cite{stefanak2}.

\begin{figure}
\begin{center}
\includegraphics[width=8.2cm]{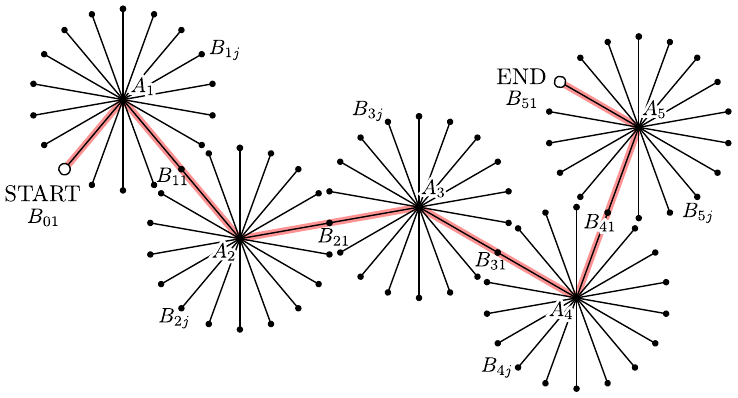}
\end{center}
\caption{In the chain of stars $\graph$ the task is to find the whole path from a known start vertex to an unknown end vertex.}
\label{fig:stars}
\end{figure}

Our aim in this paper is to examine a related task.  We consider a graph with two distinguished vertices, and we want to find the path between them. 
%In particular, we wish to find a quantum walk in which, after a certain number of steps, the particle becomes localized on the path.
In our particular case the graph $\graph$ is a bipartite graph composed of connected stars (see Fig.~\ref{fig:stars}). A star graph consists of a central vertex, which is connected to external vertices by a single edge to each external vertex, so that it looks like the hub and spokes of a wheel.  We have a string of $M$ star graphs, each having $N$ spokes, connected to each other via one of their spokes, and we do not know which vertex of star $j$ is connected to which vertex of star $j+1$, though we know the order of the stars. The first star has a vertex labeled ``START'' and the last one has a vertex labeled ``END''.  Because we do not know where the stars are connected, we do not know the path from start to end.

This task of identifying the path is reminiscent of finding one's way through a maze or movie-style safe cracking. In the latter case one must search for a single combination out of $N^M$, where $M$ is the length of the combination and $N$ is the number of settings on the dial. Clever thieves reduce this problem by looking successively for digits of the combination. This reduced task requires classically $MN/2$ choices on average. The classical situation is similar if one views the problem as finding a path through a maze. Finding the path from one distinguished vertex, ``START'', of the first star to a distinguished vertex, ``END'', on the last star then corresponds to finding the path through the maze.  Here we suppose we know what the start vertex is to simplify the discussion.

In this paper we shall show that there is a quantum walk on the graph $\graph$, for which after a number of steps proportional to the square root of the number of spokes of each star, the particle becomes localized on the path.  Then by measuring the location of the particle, we can find an element of the path.  Repeated walks and measurements can then reveal the whole path. We will start by providing an algorithm for the search with a delocalized initial state, resembling standard setups, which needs $O(M\sqrt{N}\log M)$ steps. Afterwards we will give also an algorithm that can search for the path starting from a localized initial state in $O(M\sqrt{N})$ steps.

The paper is organized as follows. In Sec.~\ref{sec:problem} we define the problem of a search for a path in a chain of stars we are aiming to solve.  Then in Sec.~\ref{sec:superposition} we present a solution based on the usual approach having a large superposition as an initial state.
To obtain a solution for a more favorable localized initial state, in Sec.~\ref{sec:locring} we solve a problem in a simplified setting on a ring of stars.
This result is then in Sec.~\ref{sec:locchain} modified to work on the chain of stars. The results are summarized in Sec.~\ref{sec:conclusion}. Some technical details are included in the Appendices.

\section{Setting of the problem}
\label{sec:problem}

Without loss of generality we can adopt the following notation for the graph, $\graph$. The $j$th star graph has central vertex $A_{j}$, and $N$ external vertices labeled $B_{j1}$ through $B_{j(N-1)}$ and $B_{(j-1)1}$.  Stars $j-1$ and $j$ share vertex $B_{(j-1)1}$, while vertex $B_{01}$ is START and vertex $B_{M1}$ is END.

For the evolution we will be using a discrete-time quantum walk formulation known as the scattering quantum walk
\cite{hillery3}.  In this walk, the particle resides on the edges of an undirected graph, and it can be
thought of as scattering when it goes through a vertex.  In particular, suppose an edge connects
vertices $v_{1}$ and $v_{2}$.  There are two states corresponding to this edge, and these states are
orthogonal --- there is the state $|v_{1},v_{2}\rangle$, which corresponds to the
particle being on the edge and going from vertex $v_{1}$ to $v_{2}$, and the state $|v_{2},v_{1}\rangle$,
which corresponds to the particle being on the same edge and going from $v_{2}$ to $v_{1}$.  The
set of these states for all of the edges forms an orthonormal basis for the Hilbert space $\hi$ of the walking particle.  

The evolution will be described by a unitary operator, $U$, that advances the walk one time step.  We obtain this
operator by combining the action of local (scattering) unitaries that describe what happens at the individual 
vertices.  In our string of stars, we have four kinds of vertices.  The simplest are $\{ B_{jk}\, |\, 1\leq j \leq M,\, 2\leq k \leq N-1\}$.  These simply reflect the particle, $U|A_{j}, B_{jk}\rangle = |B_{jk},A_{j}\rangle$.  The vertices $B_{01}$ and $B_{M1}$ also reflect the particle, but with a factor of $-1$, that is $U|A_{1},B_{01}\rangle = - |B_{01},A_{1}\rangle$ and $U|A_{M},B_{M1}\rangle = - |B_{M1},A_{M}\rangle$.  The vertices $B_{j1}$ for $1\leq j \leq M-1$ transmit the particle, i.e.\ $U|A_{j},B_{j1}\rangle = |B_{j1},A_{j+1}\rangle$ and $U|A_{j+1},B_{j1}\rangle = |B_{j1},A_{j}\rangle$.  Finally, the action of the central vertices is given for $1\leq k \leq N-1$ by
\begin{eqnarray}
U|B_{jk},A_{j}\rangle & =&  -r|A_{j},B_{jk}\rangle + t\sum_{\substack{l=1\\l\neq k}}^{N-1} |A_{j}, B_{jl}\rangle \nonumber \\
& & + t |A_{j},B_{(j-1)1}\rangle ,\label{eq:evolution1}
\end{eqnarray}
where $t=2/N$ and $r=1-t=(N-2)/N$, and
\begin{equation}
\label{eq:evolution2}
U|B_{(j-1)1},A_{j}\rangle  =  -r|A_{j},B_{(j-1)1}\rangle + t\sum_{l=1}^{N-1} |A_{j}, B_{jl}\rangle .
\end{equation}

%The unitary can be explained also in a different way. For any vertex let us set transmission coefficient $t=2/d$ and reflection coefficient $r=t-1$, with $d$ being the degree of vertex we are in. Then we can express all the local unitaries (except for START and END vertices) by this operator. Indeed, the centers of the stars act according to this description with $d=N$. For external vertices on the end of spokes $d=1$ and the scattering unitary is just one-dimensional identity operator representing pure reflection. In the case of connections between the stars $d=2$ and the local unitary is a Pauli $\sigma_x$ matrix representing pure transmission. The only difference is thus only in START and END vertices which obtain an additional minus sign; local scattering unitaries on all other vertices are described the same way.

Having defined Hilbert space $\hi$ and the evolution $U$ on this space, by choosing a proper initial state, we show how to perform an efficient quantum search for the path. We shall do it in two ways. First we will start in a large superposition of edge states and show that the problem reduces to a two-dimensional problem that is equivalent to the Grover search \cite{Grover97}. As the preparation of a complete superposition might be difficult in experimental situations, we will also investigate a case where we will choose a succession of localized initial states, which will lead to the recovery of the whole path with the same speedup.

While we use the physical interferometric analogy of the scattering approach in our paper, it is also worthwhile to comment on the possibility of an oracular setting. In computer science, search problems on graphs are formalized using a so-called oracle that, upon querying, answers a specific question. In the Grover search, the oracle answers a question, whether a queried element is the target or not. In a quantum walk setting, the oracle is more complex and presents information about the graph on which the walk takes place. In our case, the oracle that implements the scattering walk can be thought of as an operation that upon presenting a ``name'' of a vertex, outputs the names of its neighbors as well as the information as t whether it is the start or the end vertex of the searched path \cite{hillery1}.

Such an oracle encodes the path in two different ways. If the queried vertex is either the start or end, it acts as the usual oracle in the Grover search, giving direct information on whether the queried vertex is marked or not. However, the oracle encodes the connections between the stars in a different way --- when presented with the possible neighbors, the connections are recognizable by having exactly two neighbors. 

\section{Initial state of a large superposition}
\label{sec:superposition}

A calculation for two stars ($M=2$) reveals that an initial state consisting of a superposition of all of the outgoing states in the first star minus the outgoing states in the second star does lead to a state in which the particle becomes localized on the path from start to end.   The minus sign is important.  An intial state that is a superposition of all of the outgoing states in the first star plus the outgoing states in the second star leads to the particle becoming localized on the edges connected to the start and end, but provides no information about where the stars are connected.
%This suggests that for $M$ stars, an initial state that is superposition of the edge states on all of the stars, with neighboring stars having opposite signs, could produce the desired result.  Let us now show that it does.
Extrapolating from the two-star result, we start by defining the following states. % for $M$ even.
\begin{eqnarray}
|\psi_{1}\rangle & = & \frac{1}{\sqrt{M(N-2)}} \sum_{j=1}^{M}\sum_{k=2}^{N-1} (-1)^{j}|A_{j},B_{jk}\rangle,
\nonumber \\
|\psi_{2}\rangle & = & \frac{1}{\sqrt{M(N-2)}} \sum_{j=1}^{M}\sum_{k=2}^{N-1} (-1)^{j}|B_{jk},A_{j}\rangle,
\nonumber\\
|\psi_{3}\rangle & = & \frac{1}{\sqrt{2M}} \sum_{j=1}^{M} (-1)^{j}( |A_{j},B_{j1}\rangle + |A_{j},B_{(j-1)1}\rangle), \nonumber \\
|\psi_{4}\rangle & = & \frac{1}{\sqrt{2M}} \sum_{j=1}^{M} (-1)^{j}( |B_{j1},A_{j}\rangle
+ |B_{(j-1)1},A_j\rangle) .
\end{eqnarray} 
The first two states correspond to the particle being located in undesirable positions, while the next two states represent a particle being located on the path. We find that
\begin{eqnarray}
U|\psi_{1}\rangle & = & |\psi_{2}\rangle \nonumber \\
U|\psi_{2}\rangle & = & (r-t)|\psi_{1}\rangle + 2\sqrt{rt}|\psi_{3}\rangle \nonumber \\
U|\psi_{3}\rangle & = & -|\psi_{4}\rangle \nonumber \\
U|\psi_{4}\rangle & = & (t-r)|\psi_{3}\rangle + 2\sqrt{rt} |\psi_{1}\rangle .
\end{eqnarray}
so that the subspace spanned by these states is invariant under the action of $U$.  It is often the case in quantum walk search problems that the relevant states lie in an invariant subspace of small dimension \cite{krovi}.  In this case the dimension of the subspace can be reduced further by noting that
\begin{eqnarray}
U^{2}|\psi_{1}\rangle & = & (r-t)|\psi_{1}\rangle + 2\sqrt{rt}|\psi_{3}\rangle \nonumber \\
U^{2}|\psi_{3}\rangle & = & (r-t)|\psi_{3}\rangle - 2\sqrt{rt}|\psi_{1}\rangle .
\end{eqnarray}
This is already a unitary corresponding to one step of the Grover search for two elements within a database of $N$ elements. To obtain the proper initial state we continue further.
The eigenvalues of $U^{2}$ restricted to the subspace spanned by $|\psi_{1}\rangle$ and $|\psi_{3}\rangle$ are $\lambda_{\pm}=(r-t)\pm 2i\sqrt{rt}= \exp (\pm i\theta)$, for $\cos\theta=r-t$.  The corresponding eigenstates are
\begin{equation}
|\eta_{\pm}\rangle = \frac{1}{\sqrt{2}}(|\psi_{1}\rangle \mp i |\psi_{3}\rangle ) ,
\end{equation}
where $|\eta_{+}\rangle$ corresponds to $\lambda_{+}$ and $|\eta_{-}\rangle$ corresponds to $\lambda_{-}$.  We now note that
\begin{eqnarray}
U^{2n}|\psi_{1}\rangle & = & \frac{1}{\sqrt{2}}U^{2n}(|\eta_{+}\rangle + |\eta_{-}\rangle ) \nonumber \\
& = & \frac{1}{\sqrt{2}} (e^{in\theta} |\eta_{+}\rangle + e^{-in\theta} |\eta_{-}\rangle ) .
\end{eqnarray}
To localize the particle in this case in state $|\psi_3\rangle$, which is the desired effect, we look at the success probability of ending there. This turns out to be $p_{suc}(2n)=\sin^2(n\theta)$ (we emphasize the double use of the unitary explicitly in the whole paper; odd steps will be disregarded).
Choosing the number, $2n_0$, such that $n_0\theta=\pi/2$ will result in $p_{suc}(2n_0)=1$; the closest even number to $2n_0$, the number of steps we shall make, will introduce errors to the probability, which are of order $1/\sqrt{N}$, and hence will not affect our results on efficiency, which we shall present in the limits of large $N$ and $M$..

In the limit of large $N$ we obtain the number of steps (uses of $U$ or efficiency) in the quantum walk search
\begin{equation}
2n_0=\frac{\pi}{\theta}\simeq \frac{\pi}{2}\sqrt{\frac{N}{2}}.
\end{equation}
This result holds when we start from state $|\psi_1\rangle$. Using this state as the initial state  would, however, imply that we know the path. There is a state, though, that is close to $|\psi_1\rangle$, which treats all the stars' spokes equally, thus requiring no initial information about the path. This is a state that has the same amplitude for all of the outgoing edges and alternating signs on subsequent stars,
\begin{eqnarray}
|\psi_{init}\rangle & = & \frac{1}{\sqrt{MN}}\left[ \sum_{j=1}^{M}\sum_{k=1}^{N-1} (-1)^{j} |A_{j},B_{jk}\rangle \right. \nonumber \\
& & \left. + \sum_{j=1}^{M} (-1)^{j} |A_{j},B_{(j-1)1}\rangle \right] \nonumber \\
& = & \frac{1}{\sqrt{MN}}\left(\sqrt{M(N-2)}|\psi_1\rangle+\sqrt{2M}|\psi_3\rangle\right) \label{eq:superposedinit} \\
& = & \cos\frac{\theta}{2}|\psi_1\rangle + \sin\frac{\theta}{2} |\psi_3\rangle \simeq |\psi_{1}\rangle + O(N^{-1/2})|\psi_3\rangle,\nonumber
\end{eqnarray}
where the last approximation is for large $N$ and leads to a difference of $O(1/N)$ in the success probability  (for a more detailed explanation see Appendix \ref{sec:errors1}). The result is that after $2n_0$ steps (uses of the unitary $U$), the particle is localized on the path connecting start and end vertices and a measurement in the canonical (edge) basis reveals one of the connections at random from the uniform distribution (except on the edges connected to the start and end vertices, which have half the probability to be found as the edges connecting stars). As shown in Appendix \ref{sec:logsteps}, by repeating the algorithm $O(M\log M)$ times we can recover the whole path with the expected number of steps being $O(M\sqrt{N}\log M)$, obtaining a speedup over the classical case, which requires $O(MN)$ steps on average.

\section{Evolution of a localized state on a ring of stars}
\label{sec:locring}

The previous analysis contains two problems. Firstly, the efficiency $M\sqrt{N}\log M$ is not optimal and can be further improved.
Secondly, the initial state we produced in the previous section is a large superposition of edge states from the whole Hilbert space. In reality, such states are hard to create and a simpler option would be a small superposition on spatially localized edge states. Furthermore, the whole graph $\graph$ might be encoded in an oracle representing the search space and, thus, inaccessible to us.
We can draw some information from the works \cite{hein, stefanak, stefanak2}, where an analogy between searches and state transport is investigated. In similar way we shall explore the possibility of starting on one of the connections and observe ``transport'' of the state to the neighboring connections. Due to the complexity of the underlying graph the analysis is more involved, but the analogy is fitting.

We will consider an initial state localized around the start vertex that is known and subsequent initial states localized around known connections between stars. In the beginning, when only the start vertex is known, we prepare initial state
\begin{equation}
\label{eq:localizedinitstart}
|\psi_{init}\rangle=|A_1,B_{01}\rangle
\end{equation}
and if we already know the \emph{$k$-th connection} (connecting stars $k$ and $k+1$), we prepare initial state
\begin{equation}
\label{eq:localizedinit}
|\psi_{init}\rangle=\frac{1}{\sqrt{2}}(|A_{k+1},B_{k1}\rangle-|A_{k},B_{k1}\rangle).
\end{equation}
In both cases we are going to show that the evolution leads to a localization on the next connection.

Intuitively, the evolution on the star(s) we start at is roughly similar to the evolution in the Grover search, as the centers $A_{j}$ are almost reflections with phase flips and to some extent separate the two stars we start on from the rest. So we can expect that after $O(\sqrt{N})$ steps we will find the next connection, or the end vertex, depending on our starting position. The procedure of finding the whole path then requires $O(M\sqrt{N})$ steps, which improves on the case of the large superposition initial state from Eq.~(\ref{eq:superposedinit}).

The analysis of this approach is more involved than the previous case and we split it into two parts. In this section we consider a different problem, an evolution from the initial state (\ref{eq:localizedinit}) on a ring of stars, where the start vertex coincides with the end vertex. In the following section, this result will be used to solve the evolution for the original problem of the chain of stars.

We now assume the start vertex and the end vertex coincide and behave as all other connections between stars. The periodicity of the system allows us to express the eigenstates of $U^2$ using the Bloch theorem for periodic potentials in the form \cite{nagaj}
\begin{align}
|\Psi_m^\pm\rangle= \sum_{j=1}^M e^{2\pi ijm/M}\biggl( & c_{0,m}^\pm|A_j,B_{(j-1)1}\rangle\notag\\
&+\sum_{k=1}^{N-1}c_{k,m}^\pm|A_j,B_{jk}\rangle\biggr).
\end{align}
For each $m=0,1,\ldots,M-1$ there are $N-2$ eigenvectors with eigenvalue $-1$ that have zero overlap with our initial state from Eq.\ (\ref{eq:localizedinit}) and only two non-trivial eigenvectors (differentiated by the signs) having non-zero overlap with our initial state. An exception is $m=0$, for which the eigenvalue for both nontrivial eigenvectors is $1$ and only the eigenvector having $c_{0,0}=-c_{1,0}=1/\sqrt{2M}$, $c_{j,0}=0$ for $j\geq 2$ has non-zero overlap with $|\psi_{init}\rangle$; the other eigenstate with eigenvalue of $1$ is the equal superposition. For $m\neq 0$ the non-trivial eigenvalues are $\lambda_m^\pm=e^{\pm i\omega_m}$ with
\begin{equation}
\cos\omega_m=1-t[1-\cos(\phi_m)] ,
\end{equation}
and $\phi_m=2\pi m/M$, which implies that
\begin{equation}
\label{omega-m}
\omega_{m}\simeq \sqrt{2t[1-\cos(\phi_m)]}.
\end{equation}
The corresponding (normalized) eigenstates are given by
\begin{equation}
c_{0,m}^\pm=(c_{1,m}^\mp)^\ast=\frac{1}{\sqrt{2Mt(1+\cos\omega_m)}}\frac{1-\lambda_m^\pm}{\lambda_m^\pm-e^{i\phi_m}}r
\end{equation}
and
\begin{equation}
c_{m,k}^{\pm}=t/\sqrt{2Mt(1+\cos\omega_m)}
\end{equation}
for all other $2\leq k \leq N-1$.

\begin{figure}
\begin{center}
\includegraphics[width=8cm]{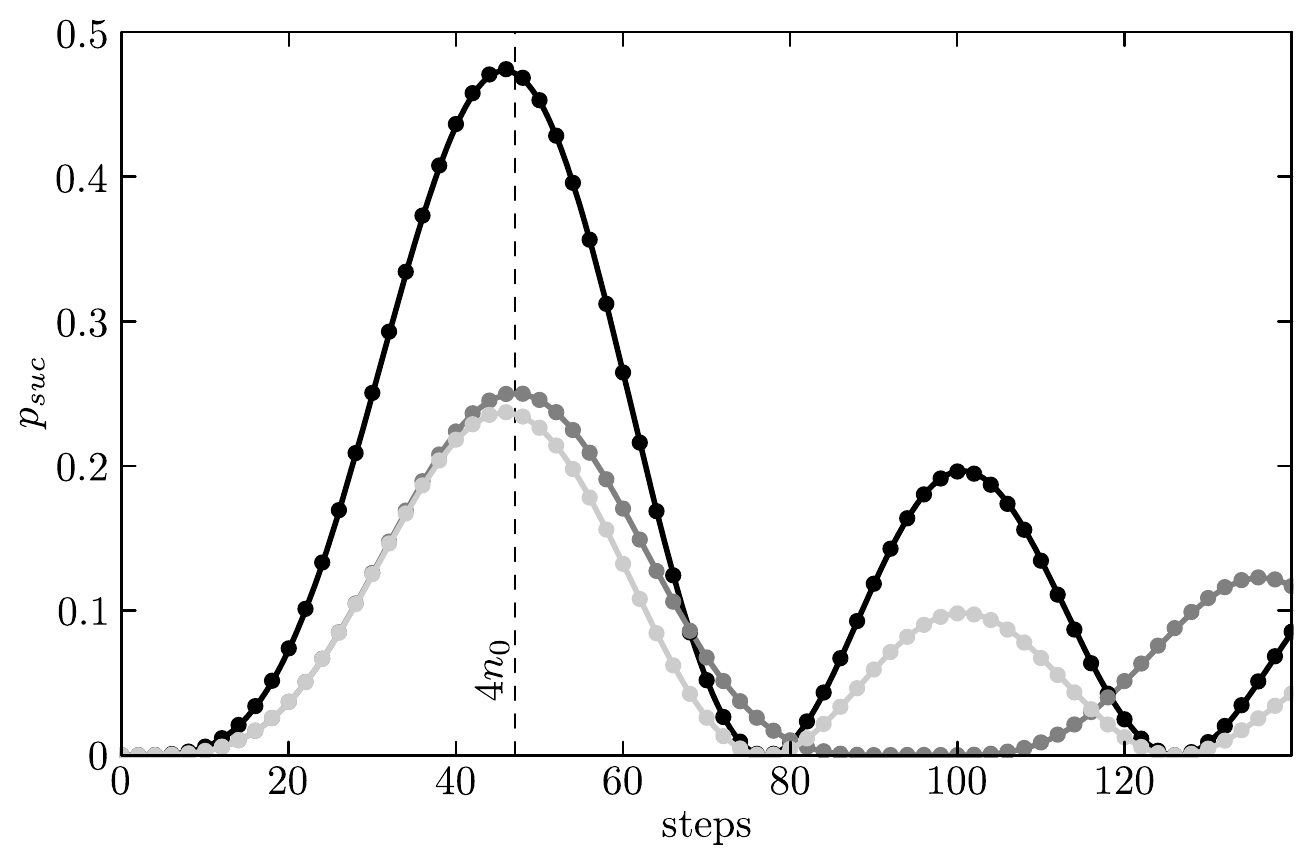}
\end{center}
\caption{\label{fig:evolutions} Typical evolutions of the success probability to find the next connection in the chain of stars for different starting stars connections with $M=11$, $N=450$---dark when starting near the start vertex, medium gray when starting at the first connection, and light when starting in the connection of some middle (fifth and sixth) stars, or without correcting terms from reflections. Dots represent exact solution from Eq.~(\ref{eq:exactprob}), while the lines are approximations by Eq.~(\ref{eq:supapprox}).}
\end{figure}

We can now expand the initial state in the eigenbasis of $U^2$, so that the state after $2n$ steps is
\begin{equation}
|\psi(2n)\rangle:=U^{2n}|\psi_{init}\rangle=\sum_{m=0,\pm}^{M-1}\langle \Psi_m^\pm|\psi_{init}\rangle e^{\pm in\omega_m}|\Psi_m^\pm\rangle.
\end{equation}
In general, let us try to find the amplitude for the state in the connection $k+b$ (we start at connection $k$). The states corresponding to the connection at $k+b$ are
\begin{align}
|e^{(b)}_+\rangle &=|A_{k+1+b},B_{(k+b)1}\rangle,\notag \\
|e^{(b)}_-\rangle &= -|A_{k+b},B_{(k+b)1}\rangle. \label{eq:successstates}
\end{align}
The corresponding amplitudes are
\begin{multline}\label{eq:exact}
E^{(b)}_\pm(2n;M) =\langle e^{(b)}_\pm|\psi(2n)\rangle\\
 =\frac{1}{\sqrt{2}M}\sum_{m=0}^{M-1}\left(1\mp t_m\frac{\sin\phi_m}{\sin\omega_m}\right)\cos(n\omega_m+b\phi_m),
\end{multline}
where $t_m=(1-\delta_{m,0})t$,
and the success probability to get located in one of the states (\ref{eq:successstates}) is  (see Fig.~\ref{fig:evolutions})
\begin{equation}\label{eq:exactprob}
p_{suc}(2n)= \left[E^{(b)}_+(2n;M)\right]^2 + \left[E^{(b)}_-(2n;M)\right]^2.
\end{equation}
The analysis in Appendix  \ref{sec:localizederror} shows that the restriction to integer steps introduces only small errors to probability, which are of order $1/\sqrt{N}$ and, hence will not affect our results on efficiency that we shall present in the limits of large $N$ and $M$.

The term proportional to $t_m$ in Eq.~(\ref{eq:exact}) is of order $1/N$ and we can set it to zero (in cases of large $N$), while at the same time we can replace the sum with an integral (taking $M\to\infty$).  Making use of Eq.~(\ref{omega-m}) and the integral representation for $J_n(z)$, the Bessel function of the first kind,
\begin{equation}
J_{n}(z)=\frac{1}{\pi} \int_{0}^{\pi} d\theta \cos (z\sin\theta - nz) 
\end{equation}
we find the approximation
\begin{equation}
\label{eq:approx}
E^{(b)}_\pm(2n;M)\simeq\frac{1}{\sqrt{2}}J_{2b}(2n\sqrt{t}).
\end{equation}
This then gives us $p_{suc}(2n)\simeq J_{2b}^2(2n\sqrt{t})$. This approximation works when $M\sqrt{N}\gg n$.

We can immediately see the first result for $b=1$, i.e.~the neighboring connection. By taking $2n\sqrt{t}=\pi$, when $J_2$ is close to its maximum, it is easy to prove that both success amplitudes are independent of both $M$ and $N$ and roughly $1/\sqrt{8}$ so that the success probability is approximately $1/4=O(1)$. Hence, by starting on the ring we will end on the next connection with probability of $1/4$ and with the same probability also on the previous connection, which, being known to us, is of no importance any more. Making $O(1)$ rounds (roughly four on average) one finds the next connection. The number of steps needed is proportional to $2n=\pi/\sqrt{t}=4n_0$. As we can now find connections successively, to find the whole path requires an expected number of $O(M\sqrt{N})$ steps, which improves the efficiency of the initial state of large superposition.

\section{Evolution of a localized state on a chain of stars} 
\label{sec:locchain}

\begin{figure*}
\begin{center}
\includegraphics[width=0.85\textwidth]{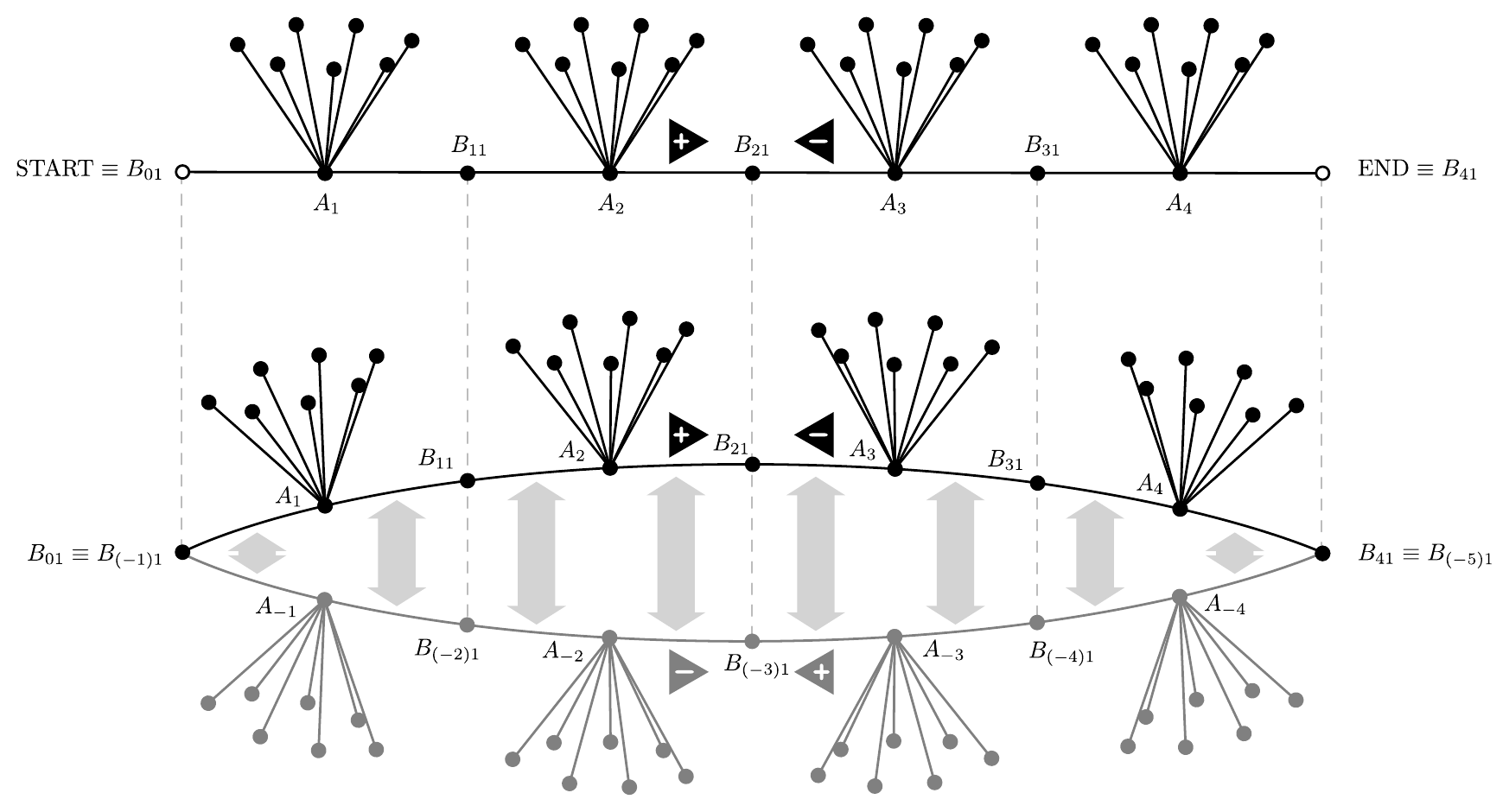}
\end{center}
\caption{\label{fig:ringchain} The correspondence between the chain of stars of length $M$ (here $M=4$) and the ring of stars of length $2M$. The state on the normal part of the ring (black) is complemented by a mirroring part (gray) of the same size but opposite sign to simulate the reflections on vertices $B_{01}$ and $B_{M1}$.}
\end{figure*}

We can use the previous result to obtain the success probabilities also on a chain of stars. The $\pi$-phased reflection on start and end vertices can be imagined as flows of oppositely signed amplitudes of the same size from some parallel chain. We will thus simulate the evolution on the chain of stars of length $M$ by introducing a ring of length $2M$. 
The correspondence is depicted in Figure~\ref{fig:ringchain}.

In the newly constructed ring graph we shall number the stars corresponding to the chain in the same way by numbers $1,2,\ldots, M$; we call this half \emph{normal}. The other half of the ring graph will be called the \emph{mirror} part, and the stars are labeled by negative numbers $-1,-2,\ldots,-M$. In this way each chain star $k$ has a mirroring counterpart $-k$. The numbering of the vertices is as follows: star centers $A_k$ and $A_{-k}$ are counterparts, as are $B_{kl}$ and $B_{(-k)l}$ for $l=2,3,\ldots,N-1$. The only inconsistencies appear in the vertices $B_{k1}$, due to the fact, that we want to simulate the evolution on the chain by evolution on the ring; these vertices are paired in the following way:
\begin{itemize}
\item vertex ``START'', previously labeled $B_{01}$, is connected to and identified with vertex $B_{(-1)1}$ of the chain,
\item vertex ``END'', labeled $B_{M1}$, is connected to and identified with vertex $B_{(-M-1)1}$ of the chain,
\item vertices $B_{k1}$ for $k=1,2,\ldots,M-1$ have their mirroring counterparts $B_{(-k-1)1}$.
\end{itemize}
The first two vertices will now act as other star connections, being purely transmissive. Having constructed a ring graph corresponding to the chain graph $\graph$ we now specify suitable states on it.

\paragraph{Mirroring states on the ring graph.} The idea behind the mapping is that now each edge state of the chain graph has its own unique ring counterpart. We will construct these ring states by pairing states, one in each part of the ring. Namely, let us define the following (un-normalized) states, which consist of normal and mirror parts, the span of which forms a subspace $\mathcal S$ of the whole ring Hilbert space:
\begin{align}
|\overline{B_{kl},A_k}\rangle := & |B_{kl},A_k\rangle - |B_{(-k)l},A_{-k}\rangle\notag\\
&\text{for $l=2,3,\ldots,N-1$,}\notag\\
|\overline{A_k,B_{kl}}\rangle := & |A_k,B_{kl}\rangle - |A_{-k},B_{(-k)l}\rangle\notag\\
&\text{for $l=2,3,\ldots,N-1$,}\notag\\
|\overline{B_{k1},A_l}\rangle := & |B_{k1},A_l\rangle - |B_{(-k-1)1},A_{-l}\rangle\notag\\
&\text{for $l\in\{k,k+1\}$,}\notag\\
|\overline{A_l,B_{k1}}\rangle := & |A_l,B_{k1}\rangle - |A_{-l},B_{(-k-1)1}\rangle\notag\\
&\text{for $l\in\{k,k+1\}$.}
\end{align}
We shall refer to these states as \emph{mirroring} states.

Acting with the unitary for the ring, $U$, on these states shows, that although these are states from the ring, the evolution, when restricted to the subspace $\mathcal S$ is the same as the evolution on a chain. In particular:
\begin{align}
U|\overline{A_k,B_{kl}}\rangle &=|\overline{B_{kl},A_k}\rangle\qquad\text{for $l\neq 1$,}\notag\\
U|\overline{A_k,B_{k1}}\rangle & =|\overline{B_{k1},A_{k+1}}\rangle\qquad\text{for $k\neq M$,}\notag\\
U|\overline{A_M,B_{M1}}\rangle & =-|\overline{B_{M1},A_{M}}\rangle,\notag\\
U|\overline{A_k,B_{(k-1)1}}\rangle &=|\overline{B_{(k-1)1},A_{k-1}}\rangle\qquad\text{for $k\neq 1$,}\notag\\
U|\overline{A_1,B_{01}}\rangle &=-|\overline{B_{01},A_{1}}\rangle.
\end{align}
The rest is described by the same equations as Eqs.~(\ref{eq:evolution1}) and (\ref{eq:evolution2}), but with the mirroring (bar) versions of states. Now we have established that the evolution of the mirroring state on the ring of stars is described by the same equations as the evolution of the original state on the chain of stars when restricted to $\mathcal S$.

\paragraph{Resulting probabilities.} Due to the symmetry of the subspace $\mathcal S$, the amplitudes on the opposing edge states in the ring are always of opposing sign and of the same size. This means, that since we did not normalize the mirroring states, the squares of amplitudes on the normal side of the ring sum up to $1$ and provide all the information about the probability distribution on the chain; i.e., we only need to measure the position of the particle only on the normal side of the ring. At the same time, the flow of the amplitude from and to the mirror part of the ring \emph{simulates} the reflections on the vertices $B_{01}$ and $B_{M1}$ of the chain.

Let us first consider an initial state (\ref{eq:localizedinitstart}) on the chain. The corresponding initial state on the mirroring ring is
\begin{equation}
|\psi_{init}\rangle=|\overline{A_1,B_{01}}\rangle = |A_1,B_{01}\rangle-|A_{-1},B_{01}\rangle.
\end{equation}
This state is just the renormalized initial state of the type of Eq.~(\ref{eq:localizedinit}) on the ring and the success amplitudes and success probabilities for the following connection ($b=1$ are simply $\sqrt{2}E_\pm^{(1)}(2n;2M)\simeq J_{2}(2n\sqrt{t})$ and $p_{suc}(n)\simeq 2J^2_{2}(2n\sqrt{t})$, respectively.

Let us now consider an initial state (\ref{eq:localizedinit}) on the chain. The corresponding initial state on the mirroring ring is
\begin{align}
|\psi_{init}\rangle = & \frac{1}{\sqrt{2}}(|\overline{A_{k+1},B_{k1}}\rangle - |\overline{A_k,B_{k1}}\rangle)\notag\\
=& \frac{1}{\sqrt{2}}(|A_{k+1},B_{k1}\rangle - |A_k,B_{k1}\rangle \notag\\
&- |A_{-k-1},B_{(-k-1)1}\rangle + |A_{-k},B_{(-k-1)1}\rangle).
\end{align}
%The first two terms correspond to the normal part of the state and the second two terms to the mirror part of the state. The amplitudes for the connection $b$ positions away from connection $k$ is then composed of two terms. The first one coming from the normal part is equal to $E_\pm^{(b)}(2n;2M)$ which was defined in Eq.~(\ref{eq:exact}).  The mirror part of the state needs to travel longer distance, the same as the part of the normal state would when rflecting from the START vertex. The mirror part going through the START vertex has to traverse the distance to the corresponding connection $B_{01}$ and then back on the normal part of the ring. The correction to the success amplitude is then given by $E_\pm^{(2k+b)}(2n;2M)$. The overall amplitudes on the desired star connections are the sums of these terms, i.e.
The first two terms correspond to the normal part of the state and the second two terms correspond to the mirror part of the state. The amplitudes for the connection $b$ positions away from connection $k$ are then composed of two terms. The first one coming from the normal part is $E_\pm^{(b)}(2n;2M)$ with $\pm$ representing the two edges of the connection on which we measure position: 
\begin{equation}
E_\pm^{(b)}(2n;2M) = \langle e^{(b)}_{\pm}|U^{2n}|\psi_{init}^{\mathrm{norm}}\rangle .
\end{equation}
Here
\begin{equation}
|\psi_{init}^{\mathrm{norm}}\rangle = \frac{1}{\sqrt{2}}(|A_{k+1},B_{k1}\rangle - |A_{k},B_{k1}\rangle) ,
\end{equation}
is the normal part of the initial state as in Eq.~(\ref{eq:localizedinit}) and $|e_\pm^{(b)}\rangle$ are defined in Eq.~(\ref{eq:successstates}).  The mirror part of the state needs to travel a longer distance. If we go through the start vertex, the mirror part has to traverse the distance to the corresponding connection $B_{01}$ and then back on the normal part of the ring. The correction to the success amplitude is then given by $E_\pm^{(2k+b)}(2n;2M)$, where
\begin{equation}
E_\pm^{(2k+b)}(2n;2M) = \langle e^{(b)}_{\pm}|U^{2n}|\psi^{\mathrm{mirror}}_{init}\rangle ,
\end{equation}
and
\begin{equation}
|\psi^{\mathrm{mirror}}_{init}\rangle =\frac{1}{\sqrt{2}}( |A_{-k},B_{(-k-1)1}\rangle - |A_{-k-1},B_{(-k-1)1}\rangle ),
\end{equation}
is the mirror part of the initial state. The overall amplitudes on the desired star connections are the sums of these terms, i.e.
\begin{equation}
F_\pm^{(b)}(2n;k;M) = E_\pm^{(b)}(2n;2M)+E_\pm^{(2k+b)}(2n;2M).
\end{equation}
The success probability for being located on the following connection on the chain ($b=1$) is then given as
\begin{equation}
p_{suc}(2n) = [F_+(2n;k;M)]^2 + [F_-(2n;k;M)]^2,
\end{equation}
where $F_\pm(2n;k;M):=F_\pm^{(1)}(2n;k;M)$. There is one additional caveat: when the measurement is performed on the last star ($k+b=M$) the success probability includes only the $F_+$ term and previous equations have to be changed accordingly.

\paragraph{Approximations.} Let us consider $b=1$ for simplicity. If we want to use an approximation for the amplitudes in terms of Bessel functions as in Eq.~(\ref{eq:approx}), we have to be careful. This approximation is based on taking $M\to\infty$, which means that it can be used to describe what is happening at the beginning and middle of the chain, but will require modifications if we want to use it near the end vertex. In particular, it means that reflections that take place at the end vertex are not taken into account.  Near the start vertex and for short times, including the optimal time $4n_0$, we have
\begin{equation}
\label{eq:supapprox}
F_\pm(2n;k;M)\simeq J_2(2n\sqrt{t})+J_{2(2k+1)}(2n\sqrt{t}).
\end{equation}
For short times of up to $4n_0$ the Bessel functions are positive, and they decrease rapidly with the increasing index. That means that in time $4n_0$ the correction increases the success probability and is observable in the vicinity of the start vertex (up to three stars away, as the numerical results show) and does not change the efficiency of the search.

Near the end vertex, when $k$ is close to $M$, we must do something else.  In particular, we can take advantage of a symmetry of ring amplitudes.  Amplitudes on a ring of length $2M$ obey
\begin{equation}
\label{eq:symmetry}
E_\pm^{(b)}(n;2M)=E_\mp^{(2M-b)}(n;2M).
\end{equation}
Using this symmetry allows us to account for the reflection from the end vertex.  In particular, we can use it to replace the second term in Eq.\ (\ref{eq:supapprox}) giving us,
\begin{align}
F_\pm(2n;k;M) &= E_\pm^{(1)}(2n;2M)+E_\pm^{(2M-2k-1)}(2n;2M)\notag\\
& \simeq J_2(2n\sqrt{t})+J_{2(2k'-1)}(2n\sqrt{t}),
\end{align}
where $k'=M-k$ is the distance from the end vertex. As $k'$ does not depend on $M$ any more, the limit $M\to\infty$ will not have an effect on it and so the term will faithfully reconstruct the reflection of the walker from the end vertex in this limit.

\paragraph{When the start vertex is unknown.}
We have assumed that the start vertex is known to us. This assumption lets us start at this vertex and uncover the path connection by connection. However, this assumption may not hold. Let us lift this condition so that all we require is the knowledge of the order of the stars; we cannot lift this condition because of the necessity of different phases on different stars in all initial states. In this case we still know which star is the first one containing the start vertex, but we do not know the location of this vertex.

Although in this case we do not have an analytical solution, numerical simulations show that we can start in other states that lead to the same speedup up to a constant; the running time of single run is now instead of $4n_0$ halved to $2n_0$. For example, we can start in a smaller superposition on the first and second star,
\begin{eqnarray}
\label{eq:twostars}
|\psi_{init}\rangle & = & \frac{1}{\sqrt{2N}}\biggl[ \sum_{k=1}^{N-1} (|A_{2},B_{2k}\rangle-|A_{1},B_{1k}\rangle)  \nonumber \\
& &  + (|A_{2},B_{11}\rangle-|A_{1},B_{01}\rangle) \biggr].
\end{eqnarray}
This state localizes with high probability on the connection between the two stars, but gets localized also on the start vertex with probability around $1/4$ and on the connection between the second and third stars with probability of roughly $1/8$. Therefore, repeating this setup several times will uncover the start vertex and the aforementioned connection. To uncover other connections, we can now continue as in the previous case in which we know the start vertex, or we can prepare a state on the next two stars the connection of which we wish to find, which is analogous to the state in Eq.~(\ref{eq:twostars}) with a complete superposition of all edge states on the stars with phases on one of the stars being $+1$ and on the other star being $-1$. Such a setup reveals both neighboring connections with probability $1/8$, unless there is no further connection in one of these direction, i.e., we are at the start or end vertices; these are obtained with probability $1/2$. All the probabilities mentioned in this paragraph depend neither on the number of spokes, $N$, nor on the number of stars, $M$.

Hence, even if the start vertex is unknown, we can uncover the whole path with the same efficiency as in the case in which it is known.

\section{Conclusion}
\label{sec:conclusion}

We have investigated a task of finding a path in a maze that was a chain of $M$ stars with $N$ spokes. The simple approach, starting in the phase-modulated equally-weighed superposition of all states, is reducible to the Grover search, which localizes the state on the whole path in $O(\sqrt{N})$ steps. Measurement then reveals a single element of this path at random. By repeating this search we can recover the whole path in  $O(M\sqrt{N}\log{M})$ steps on average.

The preparation of the highly superposed initial state for this task may be, however, difficult. We have shown that we can recover the path star by star by successive searches of $O(\sqrt{N})$ steps and improve on the overall efficiency as well. First, we start at the beginning of the path and uncover the first star connection. To reveal the next connection, with high probability we then repeat the algorithm but start in a state localized on the newly acquired connection. The whole path is then obtained in $O(M)$ successive searches leading to $O(M\sqrt{N})$ steps for the whole process.

The solution in this case cannot be obtained directly by reducing the dimensionality of the problem; an intermediate step is necessary.  We can replace the task on the chain with the task on a ring of twice the length, for which the periodicity allows us to solve the problem analytically. This result can be then used to find an exact analytical result for the successive search on the chain of stars with the efficiency of $O(M\sqrt{N})$.

\begin{acknowledgments}
D.R. was financed by SASPRO Program No.~0055/01/01 project QWIN, cofunded by the European Union and Slovak Academy of Sciences. D.R. also acknowledges support from VEGA 2/0151/15 project QWIN and APVV-14-0878 project QETWORK. M.H. and D.K. were supported by a grant from the John Templeton Foundation.
\end{acknowledgments}

\appendix

\section{Corrections to the success probability}

\subsection{Initial state of large superposition}
\label{sec:errors1}

In the case presented in Sec.~\ref{sec:superposition} after a time of $2n_0$ the success probability to find an element from the path is $p_{suc}(2n_0)=1$. But $2n_0$ is a real number in general, while the number of steps $n$ has to be an integer. Choosing integer $n$ such that $2n$ is the closest to $2n_0$ introduces an error we would like to quantify. In order to do so, let us study a more general situation, namely, having done $2n$ steps (where we allow now ``steps'' to be also real numbers), how much the success probability changes if we make $2(n+ \varepsilon)$ steps with $\varepsilon\in[-1,1]$; this interval includes also an integer number of steps closest to $n_0$.

In particular, we are interested in quantity
\begin{equation}
\label{eq:sucerror}
\Delta_\varepsilon(2n)=|p_{suc}(2n)-p_{suc}(2(n+\varepsilon))|.
\end{equation}
In our case $p_{suc}(2n)=\sin^2\left[(2n+1)\frac{\theta}{2}\right]$ with $\cos\theta=1-4/N$; we included also the effect of small overlap of the initial state with the eigenvector $|\psi_3\rangle$. After some manipulation we obtain
\begin{equation}
\label{eq:sinbound}
\Delta_\varepsilon(2n)=\left| \sin[(2n+1+\varepsilon)\theta]\sin\varepsilon\theta \right|\leq|\sin\varepsilon\theta|\leq\sin\theta,
\end{equation}
where the last inequality holds because $\theta$ is small. Substituting for $\theta$ gives
\begin{equation}
\Delta_\varepsilon(2n)\leq2\sqrt{\frac{2}{N}}.
\end{equation}
Hence, in general, the error in the success probability for time differences smaller than one step decreases as $1/\sqrt{N}$. In the optimum case, when $2n\theta=\pi$ the bound is even stricter as the first part of Eq.~(\ref{eq:sinbound}) gives
\begin{equation}
\Delta_\varepsilon(2n)\leq\sin^2\theta=\frac{8}{N}.
\end{equation}

\subsection{Localized initial state}
\label{sec:localizederror}

In the case presented in Sec.~\ref{sec:locring} we are again interested in quantity $\Delta_\varepsilon(2n)$ from Eq.~(\ref{eq:sucerror}), but now the success probability is given by Eq.~(\ref{eq:exactprob}). After some simple manipulation we get
\begin{equation}
\label{eq:Qs}
\Delta_\varepsilon(2n)\leq Q_{+,+,\varepsilon}(2n)Q_{+,-,\varepsilon}(2n)+Q_{-,+,\varepsilon}(2n)Q_{-,-,\varepsilon}(2n),
\end{equation}
where
\begin{equation}
Q_{p,\pm,\varepsilon}(2n)=\left|E_p(2n)\pm E_p(2(n+\varepsilon))\right|,
\end{equation}
and $p=\pm$ and $E_\pm(2n)$ are given by Eq.~(\ref{eq:exact}); we dropped upper index $b$ as it will have no consequences in this computation.

Using the specific form of Eq.~(\ref{eq:exact}), we can now write
\begin{multline}
Q_{p,\pm,\varepsilon}(2n)\leq\frac{1}{\sqrt{2}M}\sum_{m=0}^{M-1}\left|a_m^p\right|\cdot\bigl|\cos(n\omega_m+\phi)\\
\pm \cos((n+\varepsilon)\omega_m+\phi)\bigr|,
\end{multline}
where we used notation $\phi=b\phi_m$ for simplicity and
\begin{equation}
a_m^\pm=1\mp t_m\frac{\sin\phi_m}{\sin\omega_m}.
\end{equation}
The modulus of $a_m^\pm$ can be bounded, irrespective of the sign of the upper index and lower index $m$, as
\begin{align}
\left|a_m^\pm\right|&\leq 1+\frac{2|\sin\phi_m|}{N\sin\omega_m}=1+\sqrt{\frac{1+\cos\phi_m}{N-1+\cos\phi_m}}\notag\\
&\leq 1+\sqrt{\frac{2}{N-2}}\leq 2.
\end{align}
We can immediately obtain
\begin{equation}
Q_{\pm,+,\varepsilon}(2n)\leq\frac{1}{\sqrt{2}M}\sum_{m=0}^{M-1}4=2\sqrt{2}.
\end{equation}
For $Q_{\pm,-,\varepsilon}(2n)$ we use inequality
\begin{multline}
|\cos(n\omega_m+\phi)-\cos((n+\varepsilon)\omega_m+\phi)|\\
=2\left|\sin\left[(2n-\varepsilon)\frac{\omega_m}{2}+\phi\right]\sin\frac{\varepsilon\omega_m}{2}\right|\\
\leq 2\left|\sin\frac{\varepsilon\omega_m}{2}\right|\leq 2\sin\frac{\omega_m}{2}=2\sqrt{\frac{2}{N}},
\end{multline}
where in the last inequality we used the fact that $\omega_m$ is small. Now,
\begin{equation}
Q_{\pm,-,\varepsilon}(2n)\leq\frac{1}{\sqrt{2}M}\sum_{m=0}^{M-1}\frac{4\sqrt{2}}{\sqrt{N}}=\frac{4}{\sqrt{N}}.
\end{equation}

Putting all the partial results together into Eq.~(\ref{eq:Qs}) we finally find that
\begin{equation}
\Delta_\varepsilon(2n)\leq\frac{16}{\sqrt{N}}.
\end{equation}
So the error in the success probability that emerges when taking an integer number of steps decreases as $1/\sqrt{N}$.

\section{Recovering the path when starting in a large superposition}
\label{sec:logsteps}

The case of an initial state in a large superposition from Sec.~\ref{sec:superposition} localizes the state of the walker on the path between the start and the end vertices. A measurement then reveals one connection at random. Here we will show that the whole path can be recovered in $M\log M$ repetitions of the search, where $M$ is the number of stars.

First, let us suppose that the probability to find an unknown connection is $p$. The average number of repetitions we need to make is then
\begin{equation}
\label{eq:repetitions}
\bar r=\sum_{r=1}^\infty (1-p)^{r-1}pr=\frac{p}{[1-(1-p)]^2}=\frac{1}{p},
\end{equation}
where we used formula
\begin{equation}
\sum_{j=1}^\infty q^{j-1}j=\frac{d}{dq}\left[\sum_{j=0}^\infty q^j\right]=\frac{1}{(1-q)^2}.
\end{equation}

Now, let us analyze the situation of uncovering the whole path. Suppose we know $k$ connections already, then the probability to uncover an unknown connection is $p_k=(M-k)/M$. Then, by Eq.~(\ref{eq:repetitions}) we need
\begin{equation}
\bar r_k=\frac{1}{p_k}=\frac{M}{M-k}
\end{equation}
repetitions on average to find that connection. The overall number of repetitions is the sum of these,
\begin{equation}
\bar r=\sum_{k=0}^{M-1}\bar r_k=M\sum_{k=1}^M\frac{1}{k}.
\end{equation}
The last sum is a truncated harmonic series which can be bounded from above by an integral, which gives
\begin{equation}
\label{eq:overallrep}
\bar r\leq M+M\int_1^M\frac{1}{k}dk=M\log M+M.
\end{equation}
The number of repetitions needed to recover the whole path is then of order $M\log M$.

In our particular case the inclusion of start and end vertices poses only a slight complication. First we can consider them to form together another connection. Overall we then have $M$ connections, which we will recover in $O(M\log M)$ repetitions by Eq.~(\ref{eq:overallrep}). After this many steps (on average) we have recovered all connections and in the worst-case scenario only the start or the end vertex. The probability to find the other one is now $p=(2M)^{-1}$; this requires an additional $2M$ steps on average, which does not change the complexity.


\begin{thebibliography}{99}
\bibitem{davidovich} Y.~Aharonov, L.~Davidovich, and N.~Zagury, Phys.\ Rev.\ A {\bf 48}, 1687 (1993).
\bibitem{aharonov} D.~Aharonov, A.~Ambainis, J.~Kempe, and U.~Vazirani, STOC 01:Proceedings of 33rd annual ACM Symposium on the Theory of Computing, 50-59 (ACM, New York, 2001). 
\bibitem{reitzner} D.~Reitzner, D.~Nagaj, and V.~Bu\v{z}ek, Acta Physica Slovaka {\bf 61}, 603 (2011)
\bibitem{manoucheri} K.~Manoucheri and J.~Wang, \emph{Physical Implementations of Quantum Walks} (Springer, Heidelberg, 2014). 
\bibitem{farhi} E.~Farhi and S.~Gutmann, Phys.~Rev.~A \textbf{58,} 915 (1998).
\bibitem{AnLu09} F.M.~Andrade, M.G.E.~da Luz, Phys.~Rev.~A \textbf{80,} 052301 (2009).
\bibitem{hillery3} M.~Hillery, J.~Bergou, and E.~Feldman, Phys.\ Rev.\ A {\bf 68}, 032314 (2003).
\bibitem{shenvi}N.\ Shenvi, J.\ Kempe, and K.\ Birgitta Whaley, Phys. Rev. A {\bf 67}, 052307 (2003).
\bibitem{potocek} V.~Poto\v{c}ek, A.~G\'abris, T.~Kiss, and I.~Jex, Phys.\ Rev.\ A {\bf 79}, 012325 (2009).
\bibitem{grid1} S.\ Aaronson and A.\ Ambainis, in Proceedings of the 44th IEEE Symposium on Foundations of Computer Science (IEEE, Los Alamitos, 2003), pp. 200-209.
\bibitem{grid2} A.\ M.\ Childs and J.\ Goldstone, Phys.\ Rev.\ A {\bf 70}, 022314 (2004).
\bibitem{complete2} D.\ Reitzner, M.\ Hillery, E.\ Feldman, and V.\ Bu\v{z}ek, Phys.\ Rev.\ A {\bf 79}, 012323 (2009).
\bibitem{hillery1}M.~Hillery, D.~Reitzner, and V.~Bu\v{z}ek, Phys.\ Rev.\ A {\bf 81}, 062324 (2010).
\bibitem{kendon2} N.\ B.\ Lovett, M.\ Everitt, R.\ M.\ Heath, and V.\ Kendon, archive:1110.4366.
\bibitem{lee} J.\ Lee, Hai-Woong Lee and M.\ Hillery, Phys.\ Rev.\ A {\bf 83}, 022318 (2011).
\bibitem{feldman} E.\ Feldman, M.\ Hillery, Hai-Woong Lee, D.\ Reitzner, Hongjun Zheng, and V.\ Bu\v{z}ek, Phys.\ Rev.\ A {\bf 82}, 040301(R) (2010).
\bibitem{hillery2}M.~Hillery, Honjun Zheng, E.~Feldman, D.~Reitzner, and V.~Bu\v zek, Phys.\ Rev.\ A {\bf }85, 062325 (2012). 
\bibitem{cottrell1} S.~Cottrell and M.~Hillery, Phys.\ Rev.\ Lett.\ {\bf 112}, 030501 (2014).
\bibitem{cottrell2} S.~Cottrell, J.\ Phys.\ A {\bf 48}, 035304 (2015).
\bibitem{PeLaPoSoMoSi08}
H.B.~Perets, Y.~Lahini, F.~Pozzi, M.~Sorel, R.~Morandotti, and
  Y.~Silberberg, \newblock Phys.\ Rev.\ Lett.\ {\bf 100}, 170506 (2008).
\bibitem{ScMaScGlEnHuSc09}
H.~Schmitz, R.~Matjeschk, C.~Schneider, J.~Glueckert, M.~Enderlein, T.~Huber,
and T.~Schaetz, \newblock Phys.\ Rev.\ Lett.\ {\bf 103}, 090504 (2009).
\bibitem{KaFoChStAlMeWi09}
M.~Karski, L.~Forster, J.-M.~Choi, A.~Steffen, W.~Alt, D.~Meschede, and
 A.~Widera, \newblock Science {\bf 325}, 174--177 (2009).
\bibitem{ScCaPo09} 
A.~Schreiber, K.~N.~Cassemiro, V.~Poto\v{c}ek, A.~G\'abris, P.J.~Mosley, E.~Andersson,
I.~Jex, and Ch.~Silberhorn, Phys.\ Rev.\ Lett.\ {\bf 104}, 050502 (2010).
\bibitem{peruzzo} A.~Peruzzo, M.~Lobino, J.~C.~F.~Matthews, N.~Matsuda, A.~Politi, K.~Poulios, X.~Q.~Zhou, 
Y.~Lahini, N.~Ismail, K.~W\"{o}rhoff, Y.~Bromberg, Y.~Silberberg, M.~G.~Thompson, and J.~L.~O'Brien, 
Science {\bf 329}, 1500, (2010).
\bibitem{schreiber} A.~Schreiber, A.~G\'abris, P.~Rohde, K.~Laiho, M.~\v{S}tefa\v{n}\'ak, V.~Poto\v{c}ek, 
C.~Hamilton, I.~Jex, and Ch.~Silberhorn, Science {\bf 336}, 55 (2012).
\bibitem{hein} B.~Hein and G.~Tanner, Phys.\ Rev.\ Lett.\ {\bf  103}, 260501 (2009).
\bibitem{stefanak} M.~\v{S}tefa\v{n}\'ak and S.~Skoup\'y, Phys.\ Rev.\ A {\bf 94}, 022301 (2016).
\bibitem{stefanak2} M.~\v{S}tefa\v{n}\'ak and S.~Skoup\'y, Quantum Inf. Process. \textbf{16,} 72 (2017).
\bibitem{Grover97} L.K.~Grover, Phys.~Rev.~Lett. \textbf{79,} 325 (1997).
\bibitem{krovi} H.~Krovi and T.~A.~Brun, Phys.\ Rev.\ A {\bf 75}, 062332 (2007).
\bibitem{nagaj} M.~Kieferov\'a, D.~Nagaj, Int.~J.~Quantum Inf.~\textbf{10,} 1250025 (2012).
\end{thebibliography}
\end{document}